\documentclass[a4paper,11pt]{article}
\pdfoutput=1
\usepackage{jcappub}

\usepackage{amsmath}
\usepackage{amssymb}
\usepackage{amsfonts}
\usepackage{mathtools}

\usepackage{lineno}
\usepackage{lscape}

\usepackage[lang=english,draft]{fixme}
\fxusetheme{color}
\FXRegisterAuthor{ce}{ace}{CE}

\usepackage{subfigure}
\usepackage{graphicx}
\usepackage{color}
\usepackage{ulem}
\usepackage{bm}
\usepackage{epstopdf}
\usepackage{hyperref}
\usepackage{url}
\usepackage{booktabs}
\usepackage{mdframed}
\usepackage{fancyvrb}
\usepackage{import}
\definecolor{notecolor}{RGB}{255,72,0}

\newcommand{\GeV}[0]{\,\mathrm{GeV}}
\newcommand{\TeV}[0]{\,\mathrm{TeV}}
\definecolor{light-gray}{gray}{0.9}

\title{Antideuterons in cosmic rays: \\sources and discovery potential
}

\author[]{Johannes Herms,}
\author[]{Alejandro Ibarra,}
\author[]{Andrea Vittino,}
\author[]{Sebastian Wild}
\affiliation[]{Physik-Department T30d, Technische Universit\"at M\"unchen, James-Franck-Stra\ss e 1, D-85748 Garching, Germany}

\emailAdd{johannes.herms@tum.de}
\emailAdd{ibarra@tum.de}
\emailAdd{andrea.vittino@tum.de}
\emailAdd{sebastian.wild@ph.tum.de}

\abstract{
Antibaryons are produced in our Galaxy in collisions of high energy cosmic rays with the interstellar medium and in old supernova remnants, and possibly, in exotic sources such as primordial black hole evaporation or dark matter annihilations and decays. The search for signals from exotic sources in antiproton data is hampered by large backgrounds from spallation which, within theoretical errors, can solely account for the current data.
Due to the higher energy threshold for antideuteron production, which translates into a suppression of the low energy flux from spallations, antideuteron searches have been proposed as a probe for exotic sources.
We perform in this paper a comprehensive analysis of the antideuteron fluxes at the Earth expected from known and hypothetical sources in our Galaxy, and we calculate their maximal values consistent with current antiproton data from AMS-02.  We find that supernova remnants generate a negligible flux, whereas primordial black hole evaporation and dark matter annihilations or decays may dominate the total flux at low energies. On the other hand, we find that the detection of cosmic antideuterons would require, for the scenarios studied in this paper and assuming optimistic values of the coalescence momentum and solar modulation, an increase of the experimental sensitivity compared to ongoing and planned instruments by at least a factor of 2.
Finally, we briefly comment on the prospects for antihelium-3 detection.}

\arxivnumber{TUM-HEP 1063/16}

\begin{document}
\maketitle


\section{Introduction}

Baryons account for approximately 4.9\% of the total energy density of the Universe~\cite{Ade:2015xua} and are predominantly found in the form of hydrogen and helium-4 nuclei, along with traces of deuterium, helium-3 and lithium-7 and -6,  mainly produced during Big Bang Nucleosynthesis (for a review, see~\cite{Cyburt:2015mya}). Heavier elements, such as carbon and oxygen, can also be found in the interstellar medium (ISM) and were mostly produced in stars and subsequently ejected into space in supernova explosions (for a review, see~\cite{RevModPhys.69.995}).

Antibaryons, in contrast, are much more rare in our Universe. In fact, in the standard hot Big Bang scenario the expected relic abundance of antibaryons is negligibly small, suppressed by a factor $\sim 10^{-22}$, as the annihilations between protons ($p$) and antiprotons (${\bar p}$) persisted until the temperature of the Universe dropped below $\sim 20\,$MeV, much below the proton mass (see {\it e.g.}~\cite{Kolb:1990vq}). Nonetheless, soon after the antiproton discovery at the Bevatron~\cite{Chamberlain:1955ns}, it became clear~\cite{Fradkin} that antiprotons should be present in our Galaxy, since they should also be produced in collisions of high energy cosmic rays (CRs) with the interstellar gas, a process now dubbed secondary production (for early estimates of the galactic antiproton-to-proton ratio see~\cite{Gaisser:1973nz,Badhwar1975}). First evidence for galactic antiprotons was reported in~\cite{Golden:1979bw}, with an estimated number of events of 28.4, corresponding to an antiproton-to-proton flux ratio $\simeq(5.2\pm 1.5)\times 10^{-4}$ in the energy interval $\sim 5.6-12.5$ GeV.  Since then, experiments have steadily increased the number and the energy range of the events detected. Nowadays, the best measurements are provided by the Alpha Magnetic Spectrometer (AMS), currently in operation on the International Space Station, which has detected $3.49\times 10^5$ galactic antiproton events, permitting a precise measurement of the ${\bar p}$ flux and the $\bar p/p$ ratio in the rigidity range from 1 to 450 GV~\cite{Aguilar:2016kjl}. 

Antiproton data are, within theoretical errors, in agreement with the expectations from secondary production~\cite{Kappl:2015bqa,Evoli:2015vaa,Giesen:2015ufa}. On the other hand, the antiproton flux may also receive contributions from old supernova remnants (SNRs)~\cite{Blasi:2009bd} and, possibly, from exotic sources in our Galaxy, such as primordial black hole (PBH) evaporation~\cite{Kiraly:1981ci,Turner:1981ez} and dark matter (DM) annihilation~\cite{Silk:1984zy} or decay~\cite{Ibarra:2008qg}; production in  these exotic antiproton sources is commonly referred to as primary production.  Unfortunately, due to the large backgrounds from secondary production, it seems difficult to identify a possible contribution from exotic antibaryon sources in the current antiproton data. 

A promising avenue to discover exotic antibaryon sources in our Galaxy consists in the search for heavier antinuclei, with mass number $A\geq 2$. While the energy threshold for secondary production of one antiproton is $7m_p$, for one antideuteron (${\bar d}$) it is $17m_p$. As a result, the low energy flux of cosmic antideuterons from secondary production is expected to be very small~\cite{Chardonnet:1997dv},  thus offering an essentially background free probe of primary production mechanisms. The antideuteron flux expected at the Earth from PBH evaporation was first studied in~\cite{Barrau:2002mc}, from DM annihilation in~\cite{Donato:1999gy} and from DM decay in~\cite{Ibarra:2009tn}. Analogously, the production threshold for antihelium-3 ($^3 \overline{\rm He}$) and the unstable antitritium ($^3\overline{\rm H}$) isotope is $31 m_p$, resulting in a negligible low energy $^3 \overline{\rm He}$ flux at the Earth from secondary production; this background free probe was used in~\cite{Carlson:2014ssa, Cirelli:2014qia} to search for DM annihilations and decays. 

No cosmic antinucleus with $A\geq 2$ has yet been detected. The best current limits on the flux of cosmic antideuterons were set by the BESS collaboration, $\Phi_{\bar d}<1.9\times 10^{-4} \,{\rm m}^{-2} {\rm s}^{-1} {\rm sr}^{-1} {\rm (GeV/n)}^{-1}$ in the range of kinetic energy per nucleon $0.17\leq T\leq 1.15\,{\rm GeV/n}$ ~\cite{Fuke:2005it}, and on the cosmic antihelium-to-helium fraction by the BESS-Polar collaboration, ${\overline {\rm He}}/{\rm He}<1.0 \times 10^{-7}$ in the rigidity range from 1.6 to 14 GV~\cite{Abe:2012tz}. The sensitivity of experiments to the cosmic antideuteron flux will increase significantly in the next few years. Concretely, AMS-02 is currently searching for cosmic antideuterons in two energy windows, $0.2\leq T\leq 0.8~{\rm GeV/n}$ and $2.4\leq T\leq 4.6~{\rm GeV/n}$, with an expected flux sensitivity $\Phi_{\bar d}=2\times 10^{-6} {\rm m}^{-2} {\rm s}^{-1} {\rm sr}^{-1} {\rm (GeV/n)}^{-1}$ and $\Phi_{\bar d}=1.4\times 10^{-6} {\rm m}^{-2} {\rm s}^{-1} {\rm sr}^{-1} {\rm (GeV/n)}^{-1}$, respectively~\cite{Aramaki:2015pii}. Furthermore, the balloon borne General Antiparticle Spectrometer (GAPS) is planned to undertake a series of high altitude flights in Antarctica searching for cosmic antideuterons in the range of kinetic energy per nucleon $0.05\leq T\leq 0.25~{\rm GeV/n}$, with a sensitivity $\Phi_{\bar d}=2\times 10^{-6} {\rm m}^{-2} {\rm s}^{-1} {\rm sr}^{-1} {\rm (GeV/n)}^{-1}$~\cite{Aramaki:2015laa}. Concerning antihelium, AMS-02 is expected to improve the results of previous experiments in a wide rigidity range that extends from 1 GV to 150 GV. In particular, for a data-taking period of 18 years, its sensitivity will reach the antihelium-to-helium fractions ${\overline {\rm He}}/{\rm He} = 4 \times 10^{-10}$ and ${\overline {\rm He}}/{\rm He} = 2 \times 10^{-9}$ respectively below and above 50 GV~\cite{2010arXiv1009.5349K}. 

While the low energy ${\bar d}$ and $^3 \overline{\rm He}$ fluxes can be enhanced by additional production mechanisms, this enhancement is limited by the lack of evidence for new contributions in the antiproton flux, to which the ${\bar d}$ and $^3 \overline{\rm He}$ fluxes are highly correlated. In this paper, we calculate the maximal ${\bar d}$ and $^3 \overline{\rm He}$  fluxes for all proposed antibaryon sources in our Galaxy: secondary production, production in SNRs,  DM annihilation and decay, and PBH evaporation. Using the latest AMS-02 antiproton data~\cite{Aguilar:2016kjl} and the latest determination of the secondary antiproton production cross section from~\cite{Kappl:2014hha}, we update previous calculations on the secondary antideuteron and antihelium fluxes, and we update  their maximal fluxes from DM annihilation and decay compatible with antiproton data. Furthermore, we investigate for the first time the production of ${\bar d}$ and $^3 \overline{\rm He}$ in SNRs, and calculate their maximal fluxes. Finally,  and elaborating on previous calculations on the antideuteron flux from PBH evaporation, we calculate for the first time the maximal ${\bar d}$ and $^3 \overline{\rm He}$ fluxes from this exotic production mechanism consistent with antiproton limits.

The paper is organized as follows. In Section~\ref{sec:components} we review the various mechanisms of antinuclei production in our Galaxy: secondary production, SNRs, DM annihilation and decay, and PBH evaporation, and in Section~\ref{sec:propagation} we discuss their propagation in the Galaxy and in the heliosphere. In Section~\ref{sec:analysis} we calculate the source term of antiprotons and antideuterons for the various sources of antinuclei. We derive the maximal antideuteron flux at Earth compatible with the AMS-02 measurements of the ${\bar p}$ flux and comment on the prospects for antideuteron detection at GAPS and at AMS-02.  In Section~\ref{sec:antihe} we briefly address production of antihelium-3 in these sources and we calculate the maximal fluxes at Earth. Finally, we present in Section~\ref{sec:conclusions} our conclusions and in Appendix~\ref{sec:production} the details to calculate the antideuteron yield in the framework of the coalescence model. 

\section{Sources of antinuclei in the Galaxy}
\label{sec:components}

The flux of a given species of antinuclei at the location of the Earth is the sum of multiple contributions, which arise both from conventional astrophysical processes, as well as from possibly existing exotic sources. In the following, we present the formalism  to calculate the source spectra of antinuclei $\bar{N}$ expected from four different mechanisms: We start by discussing the production of antinuclei by spallation of CRs on the interstellar matter in Section~\ref{sec:secondary}. Then, in Section~\ref{sec:snr}, we consider the possibility that antinuclei are produced in SNRs, more specifically by spallations of the CRs directly at the source. Finally, in Sections~\ref{sec:dm} and~\ref{sec:pbh} we investigate the source spectra expected from the annihilation or the decay of DM particles and from the evaporation of PBHs.

\subsection{Antinuclei from secondary production in the interstellar medium}
\label{sec:secondary}
High energy CRs can collide with nuclei in the ISM producing secondary particles, including antinuclei. The production of antinuclei through spallation of CR protons, helium and antiprotons on interstellar hydrogen and helium has been discussed by various authors in the past, see e.g.~\cite{Duperray:2005si,Donato:2008yx,Bringmann:2006im,Ibarra:2013qt,Kappl:2014hha}. The source spectrum of antinuclei $\bar N$ with energy $E$ is given by\footnote{We use throughout the text natural units: $\hbar=c=k_B=1$.} 
\begin{align}
\label{eqn:Qsec_general}
Q^{\text{sec}}_{\bar N} \left(\vec{r}, E \right) = \sum_{i \in \left\{ p, \, \text{He}, \, \bar{p} \right\}} \sum_{j \in \left\{ \text{H}, \, \text{He} \right\}} 4 \pi \, n_j(\vec{r}) \int_{E_{\text{min}}^{\left( i,\bar{N} \right)}}^{\infty} \text{d} E_{i} \, \frac{\text{d} \sigma_{ij \rightarrow \bar N + X} \left( E_i \rightarrow  E \right)}{\text{d} E} \, \Phi_{i} \left( \vec{r}, E_{i} \right) \,.
\end{align}
Here, $\text{d} \sigma_{i,j} \left( E_i, \, E \right)/\text{d} E$ is the differential cross section for the inclusive reaction $i + j \rightarrow \bar N + X$, while $E_{\text{min}}^{\left( i,\bar{N} \right)}$ is the minimal energy of the incoming particle required for the production of the antinucleus $\bar N$ in the process.  $\Phi_{i} \left( \vec{r},  E_{i} \right)$ is the interstellar flux at the Galactic position $\vec{r}$ of the CR species $i$, while $n_\text{H}$ and $n_\text{He}$ are the densities of the interstellar hydrogen and helium nuclei.

\subsection{Antinuclei from supernova remnants}
\label{sec:snr}

SNR shockwaves are believed to be the dominant source of galactic primary cosmic rays~\cite{Blasi:2013rva}. The production and acceleration of secondary cosmic rays in SNRs has been studied in~\cite{Berezhko:2003pf} and gained attention as a possible explanation for the anomalous high-energy positron data~\cite{Blasi:2009hv,Ahlers:2009ae,Kachelriess:2011qv}.
Subsequently the production of antiprotons~\cite{Fujita:2009wk,Blasi:2009bd,Kachelriess:2011qv} and secondary nuclei~\cite{Mertsch:2009ph,Thoudam:2011aa,Tomassetti:2012ir,Cholis:2013lwa} has also been considered. In this section, we present an updated calculation of the source spectrum of antiprotons originating from SNRs, and we discuss for the first time the production of antideuterons via this mechanism. In our derivation of the SNR contribution to the antinucleon source spectrum we follow the formalism of~\cite{Tomassetti:2012ir}, which we briefly recapitulate in the following.

For the quantitative description of acceleration of CRs within the stationary plane shock model in the test particle approximation~\cite{Drury:1983zz}, it is convenient to work in a frame in which the shock front is at rest at $x=0$. Un-shocked plasma flows in from the upstream ($x<0$) region with speed $u_\mathrm{1}$ (which corresponds to the shock speed in the rest frame of the SNR star), with density $n_\mathrm{1}$ and composition of the ISM. The shocked plasma in the downstream region ($x>0$) recedes from the shock with a smaller speed $u_\mathrm{2} = u_\mathrm{1} / r$ and with an increased density $n_\mathrm{2}=r n_\mathrm{1}$, where $r$ is the compression ratio of the shock. In that reference frame, the effect of the SNR shock on the phase space density $f(x,p)$ of a primary or secondary CR species is modelled by a stationary diffusion/convection equation
\begin{equation}
u \frac{\partial f}{\partial x} = D \frac{\partial ^2 f}{\partial x^2} +\frac{1}{3} \frac{\mathrm{d}u}{\mathrm{d}x} p \frac{\partial f}{\partial p} - \Gamma^{\mathrm{inel}}f + q \,.
\label{snr-diffeqn}
\end{equation}
Here, $D$ is the diffusion coefficient, for which we adopt a Bohm form: $D = D_0 p = K_B  p / e B$, where $B$ is the strength of the magnetic field and $K_B$ is a fudge factor allowing for faster diffusion. Besides, $\Gamma^\mathrm{inel}$ is the loss rate due to inelastic scattering off the background plasma and $q$ is a source term, corresponding to thermal injection from the background plasma at the shock in the case of primary CRs, or to production in collisions of primary CRs with the background plasma in the case of secondary CRs. The source term associated to secondary production of antinuclei is given by
\begin{align}
	q_{\bar N}(x,p) = \frac{1}{4 \pi p^2} \int_0^{\infty} \mathrm{d}p_p \frac{\mathrm{d}\sigma_{pp\rightarrow \bar N +X}(p_p \rightarrow p)}{\mathrm{d}p} n \beta_p 4 \pi p_p^2 f_{p}(x,p_p),
\end{align}
where $f_{p}(x,p_p)$ is the phase space density of protons (the only primary species we consider), $\beta_p$ is the speed of the primary CRs and $n$ is the density of the background plasma. 

For efficient acceleration, the loss rate must be much smaller than the acceleration rate ($20 \Gamma D / u^2 \ll 1$) and the losses over the lifetime of the SNR should be small ($\Gamma x/ u \ll 1$). Using these conditions, the solution to the diffusion equation for secondary antinuclei in the downstream region can be approximated by:
\begin{equation}
	f_{\bar N}(x,p) \Big|_{x>0} \simeq \underbrace{f_{\bar N,0}(p) \left( 1 - \frac{\Gamma^{\text{inel}}_{x>0} x}{u_2} \right)}_\mathcal{A}
	 + \underbrace{\frac{q_{\bar N}(p) \big|_{x>0}}{u_2}x}_\mathcal{B}.
\label{snr-downstream}
\end{equation}
Here, we have split the total phase space density of antinuclei $\bar N$ into a contribution accelerated by the shock, referred to as $\mathcal{A}$-term, and a contribution arising from standard spallation in the downstream of the shock, referred to as $\mathcal{B}$-term. The latter depends on $q_{\bar N}(p) \big|_{x>0}$, i.e.~the downstream source term of antinuclei (which is independent of $x$), while the former is proportional to $f_{\bar N,0}(p) \equiv f_{\bar N}(x=0,p)$, which is given by
\begin{equation}
	f_{\bar N,0}(p) = \alpha \int_0^p \left( \frac{p'}{p} \right)^\alpha \frac{G\left(p'\right)}{u_1} e^{-\chi(p,p')} \frac{\mathrm{d}p'}{p'}.
\label{snr-f0solution}
\end{equation}
Here $\alpha = 3r/(r-1)$ is the resulting spectral index of the accelerated CRs, $\chi(p,p')$ describes losses during acceleration,
\begin{equation}
	\chi(p,p')
		\simeq \alpha (1+r^2) \frac{\Gamma_{x<0}^{\text{inel}}}{u_1^2}\left(D(p)-D(p')\right),
\end{equation}
and $G(p)$ is the sum of upstream and downstream sources from which CRs convect/diffuse into the shock:
\begin{equation}
	G(p) \simeq \frac{1}{4 \pi p^2} \int_0^{\infty} \mathrm{d}p_p \frac{\mathrm{d}\sigma_{pp\rightarrow \bar N +X}(p_p \rightarrow p)}{\mathrm{d}p} n_1 \beta_p 4 \pi p_p^2 f_{p}(0,p_p) \frac{D(p)}{u_1}
	\times \left( \frac{p_p}{p} + r^2   \right) \,.
\label{eq:SNR_Gp}
\end{equation}
In the literature, this expression is often further simplified by assuming $p/p_p = \xi \equiv \text{const}$~\cite{Blasi:2009bd,Ahlers:2009ae}. However, it has been pointed out~\cite{Kachelriess:2011qv} that while this  ``inelasticity approximation'' is well-motivated for spallation of larger nuclei, it fails in the calculation of the antinucleus source spectrum; hence, we evaluate eq.~(\ref{eq:SNR_Gp}) numerically without employing any additional approximation.

After the shock dies down, the accelerated antinuclei are free to escape into the galactic environment. The injection spectrum of antinuclei from a single SNR can be obtained from integrating the phase space density $f_{\bar N}(x,p)$, given by eq.~(\ref{snr-downstream}), over the whole downstream region, which extends from $x=0$ to $x_\mathrm{max}=u_2 \tau_\mathrm{SNR}$. Finally, the source term of antinuclei at the galactic position $\vec r$ can be readily calculated by multiplying the injection spectrum from a single SNR by the number density of SNRs in the volume element centered at $\vec r$, which corresponds to the supernova explosion rate per volume at that location,  $\mathcal{R}_\mathrm{SNR}(\vec{r})$. Under the assumption that all SNRs in our Galaxy have the same characteristics, the result reads:
\begin{equation}
	Q^\mathrm{SNR}_{\bar N}(\vec{r},E) =\mathcal{R}_\mathrm{SNR}(\vec{r}) \int_0^{x_\mathrm{max}} 4 \pi (x_\mathrm{max}-x)^2  \beta^{-1} 4 \pi p^2 f_{\bar N}(x,p)  \mathrm{d}x \,.
\label{snr-injection}
\end{equation}

\subsection{Antinuclei from dark matter annihilation/decay}
\label{sec:dm}

Multiple astronomical and cosmological observations point towards the existence of a new particle, not contained in the Standard Model of Particle Physics, which is present today in the Universe and concretely in our own Galaxy (see {\it e.g.}~\cite{Bertone:2010zza}). Some models predict that the DM could annihilate or decay producing hadrons, thus constituting a potential source of cosmic antinuclei (for reviews, see~\cite{Cirelli:2012tf,Ibarra:2013cra}). The source terms associated to the production of antinuclei in the annihilation or decay of DM particles with mass $m_{\rm DM}$ and density distribution $\rho(\vec r)$  are given by
\begin{equation}
\begin{aligned}
Q^{\mathrm {DM,ann}}_{\bar N}(\vec{r},E) &= \frac{1}{2} \left(\frac{\rho(\vec{r})}{m_{{\mathrm {DM}}}}\right)^2\sum_f \langle \sigma v \rangle_f \frac{\mathrm{d}N_f^{\bar N}}{\mathrm{d}E}\;,&\\
Q^{\mathrm {DM,dec}}_{\bar N} (\vec{r},E) &= \frac{\rho(\vec{r})}{m_{\mathrm {DM}}} \,\sum_f \Gamma_f \frac{\mathrm{d}N_f^{\bar N}}{\mathrm{d}E}\;, &
\end{aligned}
\label{eq:Q_dm}
\end{equation}
where $f$ is used as a label of the annihilation or decay channel. The quantities $\langle \sigma v \rangle_f$ and $\Gamma_f$ represent, respectively, the annihilation cross section and decay rate of DM for the channel $f$, while $\mathrm{d}N_f^{\bar N}/\mathrm{d}E$ is the energy spectrum of the antinucleus $\bar N$ produced in a single DM annihilation/decay event. 

\subsection{Antinuclei from primordial black hole evaporation}
\label{sec:pbh}

PBHs could have formed in the early Universe~\cite{1971MNRAS.152...75H,1967SvA....10..602Z} and be present in our Galaxy at present times. As predicted by Hawking, black holes radiate particles~\cite{1974Natur.248...30H,1975CMaPh..43..199H} and  therefore may constitute an exotic source of antibaryons~\cite{Kiraly:1981ci,Turner:1981ez,Barrau:2001ev,Barrau:2002mc}. The differential emission rate per degree of freedom of the particle species $j$, with mass $m_j$ and spin $s_j$, from an uncharged, non-rotating black hole with mass $M_{\rm PBH}$ is given by~\cite{1975CMaPh..43..199H,PhysRevD.13.198}
\begin{equation}
	\frac{\mathrm{d}^2N_j}{\mathrm{d}Q \mathrm{d}t}
	= \frac{\Gamma_j}{2 \pi  \left( \mathrm{exp}\left(Q/T\right) - \left(-1\right)^{2s_j} \right)} \,.
\label{Eq:PBH_parton_mult}	
\end{equation}
Here, $Q$ is the energy of the emitted particle and $T$ is the Hawking temperature:
\begin{equation}
  T = \frac{M^2_{\rm Pl}}{ 8 \pi M_{\rm PBH}} \approx \frac{1.06 \GeV}{M_{\rm PBH} / 10^{13}\,\mathrm{g}} \,,
\end{equation}
with $M_{\rm Pl}$ the Planck mass, and $\Gamma_j$ the probability that the particle is absorbed by the black hole, which is related to the absorption cross section $\sigma_j$ by~\cite{PhysRevD.41.3052}
\begin{equation}
\Gamma_j (M_{\mathrm{PBH}},Q,m_j) = \frac{1}{\pi}\,(Q^2-m_j^2)\, \sigma_j(M_{\mathrm{PBH}},Q,m_j)\;.
\end{equation}

As pointed out in~\cite{PhysRevD.41.3052}, black holes do not directly emit antibaryons. Instead, the antibaryon production proceeds via the emission of quarks and gluons which subsequently hadronize producing antiprotons and antideuterons. The number of antiprotons or antideuterons with energy $E$ that are produced in the hadronization of a parton $j$ with energy $Q$ is given by the fragmentation function $\mathrm{d}g_{\bar{N},j}(Q,E)/\mathrm{d}E$. Then, the number of antinuclei $\bar N$ emitted per unit time and unit energy by a single black hole with mass $M_{\rm PBH}$ is:
\begin{equation}
\frac{\mathrm{d}^2 N_{\bar{N}}}{\mathrm{d}E\mathrm{d}t} = \sum_{j=q,g} \int_{Q=E}^\infty \alpha_j \frac{\mathrm{d^2}N_j 
\left( M_\mathrm{PBH} \right)
}{\mathrm{d}Q \mathrm{d}t} \; \frac{\mathrm{d}g_{\bar{N},j}(Q,E)}{\mathrm{d}E} \, \mathrm{d}Q \,,
\label{Eq:PBH_hadron_mult}
\end{equation}
where $\alpha_j$ is the number of degrees of freedom of the parton $j$.

PBHs are present in our Galaxy with a mass spectrum $\mathrm{d}{\cal N}/\mathrm{d}M_{\rm PBH}$ and a mass density distribution $\rho_{\rm PBH}(\vec r)$. Therefore, the number density distribution is $\rho_{\rm PBH}(\vec r)/\overline{M}_{\rm PBH}$, where $\overline{M}_{\rm PBH}$ is the average PBH mass, defined as:
\begin{equation}
\overline{M}_{\rm PBH}=\int \mathrm{d}M_{\mathrm{PBH}} \, \frac{\mathrm{d}{\cal N}}{\mathrm{d}M_{\mathrm{PBH}}} M_\mathrm{PBH}\;.
\end{equation}

The antinuclei source term can then be readily calculated by integrating contributions from PBHs with all masses in the volume element centered at the Galactic position $\vec r$. The result is~\cite{Barrau:2001ev,Barrau:2002mc}:
\begin{equation}
Q^{\mathrm{PBH}}_{\bar{N}} (\vec{r},E) = \int \mathrm{d}M_{\mathrm{PBH}} \frac{\mathrm{d}^2 N_{\bar{N}}}{\mathrm{d}E\mathrm{d}t} \frac{\mathrm{d}{\cal N}}{\mathrm{d}M_{\mathrm{PBH}}} \;
\frac{\rho_\mathrm{PBH}\left( \vec{r} \right)}{\overline{M}_{\rm PBH}}\;.
\label{Eq:PBH_source_term}
\end{equation}

\section{Propagation in the Galaxy and solar modulation}
\label{sec:propagation}

After being produced, charged CRs propagate across the Galaxy and interact with the interstellar gas and the galactic magnetic field.
The number density $\psi_{\bar N}$ of a CR species $\bar N$ is modelled by the following transport equation (assuming the steady-state condition holds)~\cite{1964ocr..book.....G}: 

\begin{equation}
\nabla(-D\nabla \psi_{\bar N} + \vec{V_c} \psi_{\bar N})+\frac{\partial}{\partial E} \left(b_{\mathrm {tot}} \psi_{\bar N} - K_{E}\frac{\partial}{\partial E}\psi_{\bar N} \right) = Q_{\bar N} - \Gamma_{\mathrm{ann}}\psi_{\bar N} \;,
\label{eq:transport}
\end{equation}
where $E$ is the total energy of the particle. On the left hand side of the transport equation, $D$ is the coefficient associated to spatial diffusion, which is a process caused by the interaction of the CRs with the inhomogeneities of the galactic magnetic field. We assume diffusion to be homogenous and isotropic, to be confined within a cylinder of radius $R$ = 20 kpc and half-height $L$ centered at the galactic plane and to depend only on the rigidity $\mathcal{R} = p/eZ$: 
\begin{equation}
D(\mathcal{R}) = D_0 \beta \left( \frac{\mathcal{R}}{1\,\mathrm{GV}}\right)^\delta \,.
\end{equation}
The interaction with the magnetic field of the Galaxy also generates diffusion in momentum space, known as reacceleration, whose strength is determined by the coefficient
\begin{equation}
K_E(E) = \frac{4}{3 \delta (4-\delta^2) (4 - \delta)} V_a^2 \frac{E^2 \beta^4}{D(E)},
\end{equation}
where $V_a$ is the Alfv\'en velocity of the turbulences in the magneto-hydrodynamic plasma.

The quantity $\vec{V_c} = \mathrm{sign}(z) v_c \vec{e}_z$ denotes the velocity of the convective wind, taken constant and directed out of the galactic plane.
The energy loss term $b_{\mathrm{tot}} = \text{d}E/\text{d}t$ takes into account the energy losses due to ionization and Coulomb interactions with the particles of the ISM, the adiabatic loss term related to the convective wind and the energy shift related to reacceleration (which, in principle, can also be an energy gain). For all these terms we have used the formulae given in Ref.~\cite{Maurin:2002ua}.  

On the right hand side of the transport equation, the quantity $Q_{\bar N}$ represents the source term for the species $\bar N$; the source terms for the processes relevant to this work have been discussed in the previous section.
The term $\Gamma_{\mathrm{ann}}\psi_{\bar N}$ describes losses due to annihilation and destructive inelastic scattering of CRs with the ISM:
\begin{equation}
\Gamma_{\mathrm{ann}} = \sum_{j \in \mathrm{ISM}} n_j \sigma^{\mathrm{ann}}_{\bar{N}j} v_{\bar N} \;,
\end{equation} 
where $\sigma^{\mathrm{ann}}_{\bar{N}j} $ is the total annihilation cross section for the collision of the particle $\bar{N}$ off the target particle species $j$ of the ISM, whose number density is $n_j$.
In addition to the source term and the destruction term, we also take into account the so-called tertiary contribution, which models the redistribution of particles
to lower energies as a consequence of inelastic but non-destructive interactions with the ISM. The source term for this contribution is: 
\begin{equation}
Q^{\mathrm{ter}}_{\bar{N}} (\vec{r},E) = \sum_{j \in \mathrm{ISM}} n_j(\vec{r}) \left [ \int_E^\infty \frac {\text{d}\sigma^{\mathrm{ina}}_{\bar{N} j \rightarrow \bar{N} X}}{\text{d}E}(E' \rightarrow E)\, \beta ' \psi_{\bar N} (E') dE' - \sigma^{\mathrm{ina}}_{\bar{N}j \rightarrow \bar{N}X}(E)\, \beta \psi_{\bar{N}} (E) \right]\;.
\label{Eq:Q_ter}
\end{equation}
Here $\text{d}\sigma^{\mathrm{ina}}_{\bar{N} j  \rightarrow \bar{N} X} (E' \rightarrow E)/\text{d}E$ is the differential scattering cross section for inelastic collisions of particles of species $\bar{N}$ with energy $E'$ off a target particle of species $j$ from the ISM, where the projectile $\bar{N}$ survives the interaction with a lower energy $E$. For antiprotons this corresponds to ``non-annihilating rescattering''; for heavier nuclei, the interaction has to be non-disintegrating.

We solve the transport equation within the semi-analytical framework of the two-zone diffusion model, which is described extensively in~\cite{Maurin:2002ua,Maurin:2001sj,Donato:2001eq,Maurin:2002hw,Donato:2001ms}. This propagation model is uniquely defined by the 5 parameters $D_0$, $\delta$, $V_a$, $v_c$. and $L$. We choose to fix these parameters to the values that have been found in Ref.~\cite{Kappl:2015bqa} to give the best-fit to the recent AMS-02 B/C data: $D_0=0.0967$ kpc$^2$ Myr$^{-1}$, $\delta=0.408$,  $V_a=31.9$ km s$^{-1}$,  $v_c=0.2$ km s$^{-1}$ and $L=13.7$ kpc. We expect the choice of the propagation parameters to have only a very mild effect on the  maximally allowed antideuteron signal from a given source, compatible with the constraints from the $\bar p$ data. In fact, as investigated in~\cite{Ibarra:2012cc,Fornengo:2013osa} a set of propagation parameters that enhances the antideuteron flux will also enhance the antiproton flux and this will translate into stronger limits on the parameter determining the strength of the antideuteron signal.

After their propagation across the Galaxy, CRs enter the solar system, where they interact with the magnetic field of the Sun. This solar modulation of CR fluxes can be treated analytically within the force field approximation~\cite{1968CaJPS..46..937G,1968ApJ...154.1011G,1971Ap&SS..11..288G}.
In this framework, the top-of-atmosphere flux ({\it i.e.}~the flux after modulation) is related to the interstellar flux ({\it i.e.}~the flux before modulation) by the following relation:
\begin{equation}
\Phi_{\mathrm{TOA}}(T_{\mathrm{TOA}}) = \frac{A^2 \, T_{\mathrm{TOA}}^2 + 2m_{\bar{N}}A \,T_{\mathrm{TOA}}}{A^2 \,T_{\mathrm{IS}}^2 + 2m_{\bar{N}}A\, T_{\mathrm{IS}}} \Phi_{\mathrm{IS}}(T_{\mathrm{IS}})  \;,
\end{equation}
where $T_{\mathrm{TOA}}$ and $T_{\mathrm{IS}}$ are, respectively, the kinetic energy per nucleon at the top of the Earth's atmosphere and at the heliospheric boundary, which are related by $T_{\mathrm{TOA}} = T_{\mathrm{IS}} - e \varphi |Z|/A$. Here,  $A$ and $Z$ are the mass and atomic numbers of the antinucleus, and $\varphi$ is the so-called Fisk potential. The values we assign to this parameter in our analysis are discussed in the next section. 

\section{Analysis and results}
\label{sec:analysis}

Our goal in this paper is to evaluate the largest antideuteron flux compatible with the experimental limits on antiproton production in the Galaxy. To this end, we first describe in Section~\ref{subsec:source_terms} the details of the numerical calculation of the source terms introduced in Section~\ref{sec:components}, and then, in Section~\ref{subsec:method} we present our approach to set upper limits on the free parameters of each scenario from the current AMS-02 antiproton data, and to calculate the corresponding maximal antideuteron fluxes. 

\subsection{Determination of the source terms}
\label{subsec:source_terms}

\begin{figure}
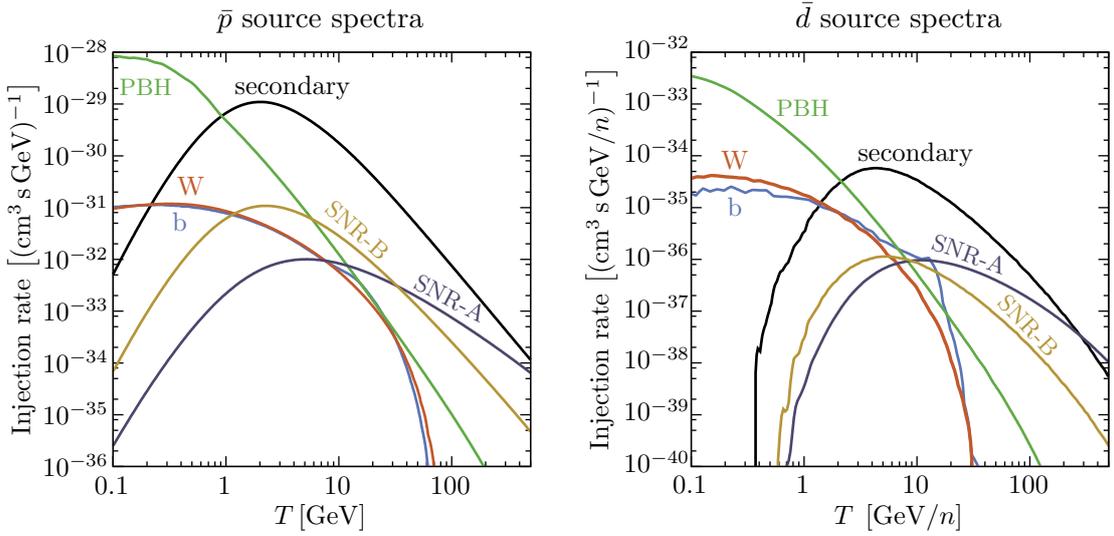

	\begin{center}
		\subimport{graphics/sourceSpectra/}{sourcePbar.tex}
		\subimport{graphics/sourceSpectra/}{sourceDbar.tex}
		\caption{Antiproton (left panel) and antideuteron (right panel) source spectra at the position of the Sun for the exemplary scenarios discussed in the text.}
		\label{fig-sources}
	\end{center}
\end{figure}

\subsubsection{Secondary production}
\label{subsubsec:secondary}

The source term associated to production of secondary antinuclei in the ISM is defined in eq.~(\ref{eqn:Qsec_general}).  To calculate the secondary antiproton flux, we consider collisions of energetic protons and helium nuclei on the ISM, which we assume to be mostly constituted by hydrogen and helium atoms. We use as incident proton and helium fluxes, $ \Phi_{\text{H}} $ and $ \Phi_{\text{He}}, $  the parametrizations presented in~\cite{Kappl:2015bqa}, which are based on AMS-02~\cite{Aguilar:2015ooa,ams_haino_talk} and CREAM~\cite{Yoon:2011aa} data, demodulated with $\varphi = 0.57\,$GV, and which we assume to be uniform throughout the Galaxy. For the antiproton production cross section, $\text{d} \sigma_{ij \rightarrow \bar p + X} \left( E_i \rightarrow  E \right)/\text{d} E$, we use the parametrization presented in~\cite{Kappl:2014hha}, which is based on the NA49 data on $p p \rightarrow \bar p X$ scattering~\cite{Anticic:2009wd}, and which takes into account possible isospin-breaking effects on antineutron production, as well as the production of antiprotons via hyperon decay. The channels involving helium are also modeled the same way as in~\cite{Kappl:2014hha}. Finally, the interstellar matter densities are taken to be $n_{\text{H}} = 0.9\,\text{cm}^{-3}$ and $n_\text{He} = 0.1\,\text{cm}^{-3}$, following~\cite{Ferriere:2001rg}.

For the calculation of the secondary antideuteron flux we also include, following~\cite{Duperray:2005si}, the contribution from collisions of energetic antiprotons on  nuclei of the ISM. For the antideuteron production cross section, $\text{d} \sigma_{ij \rightarrow \bar d + X} \left( E_i \rightarrow  E \right)/\text{d} E$, we employ the results from~\cite{Ibarra:2012cc}, which were obtained in the framework of the event-by-event coalescence model using a modified version of the Monte Carlo generator DPMJET-III (see appendix~\ref{sec:production} for details). The antideuteron yields from spallation reactions involving helium are obtained, following~\cite{Kappl:2014hha}, from rescaling the Monte Carlo results of the $pp$ and $\bar p p$ channel by appropriate nuclear enhancement factors. The resulting antiproton and antideuteron source terms are shown in Fig.~\ref{fig-sources} as black lines. 

\subsubsection{Supernova remnants}

We calculate the antinuclei source terms using eq.~(\ref{snr-injection}), which depends on various free parameters. The shape of the spectrum of antinuclei is determined by the compression ratio $r$ and by the cut-off momentum $p_\text{cut}$. We adopt for our analysis the value  $r=3.2$ in order to match the spectral index of CR protons and gamma ray sources~\cite{Ahlers:2009ae}, while we consider several values for  the cut-off momentum, $p_\mathrm{cut}=$1, 5, 20, 100 TeV. The other SNR parameters of the acceleration model described in Section~\ref{sec:snr}, namely $\tau_\mathrm{SNR},\, u_1,\, n_1$ and $K_B/B$, as well as the total supernova explosion rate in the Milky Way,  $\mathcal{R}_\text{SNR,tot}$, only affect the normalization of the fluxes (up to loss effects $<5\%$). More precisely, in the limit of small losses,  the parameter dependence of the proton yield $\mathcal{P}$ and the yields of the secondary species are
\begin{equation}
\begin{aligned}
\mathcal{P} &\propto \mathcal{R}_\mathrm{SNR,tot} Y_\mathrm{proton} \tau_\mathrm{SNR}^3 u_1^3 \\
\mathcal{A} &\propto \mathcal{P} \cdot u_1^{-2} n_1 (K_B/B) &:=& \mathcal{P} \cdot F_\mathcal{A} \\
\mathcal{B} &\propto \mathcal{P}\cdot \tau_\mathrm{SNR} n_1 &:=& \mathcal{P}\cdot F_\mathcal{B} \,,
\end{aligned}
\end{equation}
where $Y_\mathrm{proton}$ is related to the injection of thermal protons into the acceleration process. Different normalizations for $\mathcal{A}$, $\mathcal{B}$ are obtained from different sets of parameters, adjusting $\mathcal{R}_\mathrm{SNR,tot} Y_\mathrm{proton}$ to match the proton yield to the observed flux. The normalizations $N_\mathcal{A}$ and $N_\mathcal{B}$ are then defined as normalization relative to a  benchmark model, $N_{\mathcal{A},\mathcal{B}} =  F_{\mathcal{A},\mathcal{B}}/F_{\mathcal{A},\mathcal{B}}^{\mathrm{benchmark}}$. We adopt as benchmark the set of parameters $n_1 = 2\,\mathrm{cm}^{-3}$, $u_1 = 5\times 10^{7} \mathrm{cm}/\mathrm{s}$, $B/{K_B} = 1/20 \,\mu\mathrm{G}$ and $\tau_\mathrm{SNR} = 20 \,\mathrm{kyr}$, which has been widely used in recent literature in the framework of SNRs as the possible origin of the positron excess~\cite{Blasi:2009hv}. We take into account the radial distribution $\mathcal{R}_\text{SNR}(r)$ of SNRs in the Galaxy \cite{Green:2015isa}.

We show in Fig.~\ref{fig-sources} as purple (dark yellow) lines the antiproton and antideuteron source terms from SNRs for the case $N_\mathcal{A}$ = $N_\mathcal{B}$ = 1, considering only the $\mathcal{A}$-term ($\mathcal{B}$-term), for a cut-off momentum $p_\mathrm{cut} = 20 \;\mathrm{TeV}$. As seen in the plot, the energy spectrum of the source terms from spallations in the ISM and from SNRs is very similar, which is due to the fact that in both processes the antinuclei are of secondary origin. As a result, the shape of the low energy tail of the total antideuteron flux will not be significantly altered by the antinuclei production in SNRs, thus making this source difficult to disentangle from the secondary production.

\subsubsection{Dark matter annihilation/decay}

We compute the DM source term using eq.~(\ref{eq:Q_dm}). The antiproton and antideuteron energy spectra $\text{d}N_f^{\bar{p},\bar{d}}/\text{d}E$ are obtained with the Monte Carlo event generator PYTHIA 8.176~\cite{Sjostrand:2007gs}, employing the event-by-event coalescence model for the case of antideuteron production, as discussed in Appendix~\ref{sec:production}. For the annihilation/decay channels, we consider the two representative cases $f = b \bar{b}$ and $f = W^+W^-$. We assume the DM density profile $\rho(\vec{r})$ to follow a NFW distribution~\cite{Navarro:1996gj} $\rho(\vec{r}) \propto (r/r_s)^{-1} (1+r/r_s)^{-2}$, with a scale radius $r_s = 24.42\,$kpc, normalized to give a local DM mass density $\rho_0 = 0.4\,\text{GeV}/\text{cm}^3$.

In Fig.~\ref{fig-sources} we show, for illustration, the antiproton and antideuteron source terms  from DM annihilations into $b\bar{b}$ (blue lines) or $W^+W^-$ (red lines) assuming $m_{\mathrm{DM}}$ = 100 GeV. The normalization corresponds to a cross section $\langle \sigma v \rangle = 3\times 10^{-26}$ cm$^3$s$^{-1}$, as suggested by frameworks of DM production via thermal freeze-out. 

\subsubsection{Primordial black holes}

The source term associated to the PBH contribution is defined in eq.~(\ref{Eq:PBH_source_term}). We calculate the antiproton production rate from a PBH with mass $M_{\rm PBH}$ evaluating the fragmentation functions $\mathrm{d}g_{\bar{N},j}(Q,E)/\mathrm{d}E$ that enter in eq.~(\ref{Eq:PBH_hadron_mult}) by using  the PYTHIA event generator\footnote{More technically, we simulate the shower and the hadronization for a particle-antiparticle jet and then evaluate the fragmentation function as 
\begin{equation*}
\frac{\mathrm{d}g_{\bar{N},j}(Q,E)}{\mathrm{d}E} = \frac{1}{2} \,\frac{\mathrm{d}g_{\bar{N},j\bar{j}}(2Q,E)}{\mathrm{d}E}  \, ,
\end{equation*}
where $j$ is the parton under consideration.
}. For the antideuteron production rate, we employ the event-by-event coalescence model discussed in Appendix~\ref{sec:production}.  On the other hand, for the absorption cross section $\sigma_f(M_{\mathrm{PBH}},Q,m_f)$ we use the estimate derived in~\cite{PhysRevD.13.198} for massless particles (using this result for massive particles introduces an error which is negligible for $Q \gg m_f$, as is always the case for antideuteron production from light quarks and gluons, and is at most  50\% when $Q = m_f$~\cite{PhysRevD.41.3052}).

To determine the PBH mass spectrum, $\mathrm{d}{\cal N}/\mathrm{d}M_{\mathrm{PBH}}$, we assume that all PBHs today were formed from the collapse of scale invariant primordial density fluctuations~\cite{1974MNRAS.168..399C}.\footnote{ Other production mechanisms that can contribute to the PBH density today are the collapse of cosmic string loops~\cite{1989PhLB..231..237H} or through bubble collisions~\cite{1982Natur.298..538C, PhysRevD.26.2681}. For reviews, see~\cite{Carr:2005zd,Khlopov:2008qy,Green:2014faa}.} The initial mass spectrum in this case takes the form~\cite{carr_primordial_1975}
\begin{equation}
\frac{\mathrm{d}{\cal N}}{\mathrm{d}M_{\mathrm{PBH}}}\Big|_{t=0} \propto M_{\mathrm{PBH}}^{-5/2}.
\label{Eq:PBH_ini_spectrum}
\end{equation} 
The evaporation reduces the mass of the PBH at a rate
\begin{equation}
\frac{\mathrm{d}M_{\mathrm{PBH}}}{\mathrm{d}t} = -\alpha M_{\mathrm{PBH}}^{-2}, 
\end{equation}
and modifies the mass spectrum.  For the temperatures relevant for antinuclei formation in PBH evaporation, $T\gtrsim \Lambda_{\rm QCD}$, the parameter $\alpha$ can be taken approximately constant~\cite{Carr:2016hva}. Therefore, the mass spectrum today reads:
\begin{equation}
\frac{\mathrm{d}{\cal N}}{\mathrm{d}M_{\mathrm{PBH}}}\Big|_{\rm today} = A \left(\frac{{M_{\mathrm{PBH}}^*}^3}{M_{\mathrm{PBH}}^{4/3}} + M_{\mathrm{PBH}}^{5/3}\right)^{-3/2} \;,
\label{Eq:PBH_mass_spectrum}
\end{equation}
where $A$ is a normalization factor and $M_{\mathrm{PBH}}^*=\left(3\alpha t_{\mathrm{univ}}\right)^{1/3} = 5 \times 10^{14} \,\mathrm{g}$~\cite{Carr:2016hva} is the mass at the formation time of the PBHs evaporating today, which corresponds to a temperature $T(M_{\mathrm{PBH}}^*) \approx 21\;\mathrm{MeV}$. This temperature is below $\Lambda_{\rm QCD}$, therefore the mass of the evaporating PBHs producing antinuclei is much smaller than $M_{\mathrm{PBH}}^*$. As a result, for the mass range relevant for our study, the mass spectrum today approximately scales as $M_{\mathrm{PBH}}^2$. As pointed out in~\cite{Barrau:2001ev}, the fact that the mass spectrum increases as $M_\mathrm{PBH}^2$ for PBH masses much smaller than  $M_{\mathrm{PBH}}^*$ does not depend on any assumption about the initial mass spectrum, as long as the latter is nonzero and without abrupt changes around $M^*_\mathrm{PBH}$. Hence, our conclusions for the maximal antideuteron fluxes are fairly insensitive to the form of the initial mass spectrum.

Finally, and as PBHs constitute a form of cold DM, we assume that they are distributed in the Milky Way following the NFW profile, with a local mass density which is left as a free parameter. We show in Fig.~\ref{fig-sources} as green lines, the antiproton and antideuteron source terms from PBH evaporation, assuming for concreteness a local PBH mass density $\rho_{\mathrm{PBH,\odot}}=10^{-32}\,\mathrm{g}/\mathrm{cm}^3$.


\subsection{Limits from antiprotons and maximal antideuteron fluxes}
\label{subsec:method}

Among the four sources of cosmic antinuclei discussed above, secondary production plays a special role, as it is the one with the highest predictive power. Notably, this term alone reproduces, within theoretical uncertainties, the AMS-02 antiproton data~\cite{Kappl:2015bqa,Evoli:2015vaa,Giesen:2015ufa}. Therefore, we will make the reasonable assumption that the dominant contribution to the cosmic antiproton flux originates from spallations of CRs in the ISM, while the three other antinuclei production mechanisms  (SNRs, DM annihilation/decay and PBH evaporation) give only a sub-dominant contribution. 

We constrain the free parameters of each of these additional antimatter sources by means of a profile likelihood approach in which the contributions from SNRs, DM and PBHs are treated as signals, to be added to a background given by the secondary term. To each signal we associate a chi-square that is defined as follows: 
\begin{equation}
\chi^2 = \sum_i \frac{(R_{i,\mathrm{th}}(\theta, N_{\mathrm{sec}}) - R_{i,\mathrm{exp}})^2}{\sigma^2_{i,\mathrm{exp}}} \,,
\label{eq:chi_sq}
\end{equation}
where $R_{i,\mathrm{th}}$, $R_{i,\mathrm{exp}}$ and $\sigma_{i,\mathrm{exp}}$ are, respectively, the theoretical prediction, the experimental measurement and the one-sigma error of the measurement in the $i$-th energy bin~\cite{Aguilar:2016kjl}. We find that the background-only hypothesis gives $\chi^2/{\rm d.o.f}=2.1$ for the antiproton flux and $\chi^2/{\rm d.o.f}=4.4$ for the antiproton-to-proton ratio. We will therefore use the antiproton flux to probe other possible antinuclei production mechanisms in our Galaxy.

The quantity $R_{i,\mathrm{th}}$ depends on the value of the two parameters $N_{\mathrm{sec}}$ and $\theta$. The former is the normalization of the antiproton secondary flux, which we allow to vary between a factor of 0.9 and 1.3 with respect to our nominal result, in order to take into account the uncertainty on the antiproton spallation cross section (see \cite{Kappl:2014hha}). The latter corresponds to the parameter that determines the amplitude of the antiproton flux for the different sources under consideration: $\theta = \langle \sigma v \rangle$ ($\Gamma$) for DM annihilation (decay), $\theta = N_{\mathcal{A}}, N_{\mathcal{B}}$ for SNR $\mathcal{A}$ and $\mathcal{B}$ terms, and  $\theta = \rho_\mathrm{PBH} $ for PBHs.

We use the chi-square defined in eq.~(\ref{eq:chi_sq}) to derive $95\%$ C.L.~upper limits on the amplitude of each signal, treating $N_{\mathrm{sec}}$ as a nuisance parameter. The limit corresponds to the value of $\theta$ which allows the largest signal compatible with the condition $\chi^2 \leq \chi^2_\mathrm{min} + 4$, 
where the minimum of the chi-square $\chi^2_\mathrm{min}$ is determined for each signal independently. In order to avoid degeneracies with the signal amplitude at low energies, we fix the Fisk potential to $\varphi = 0.90 \,\mathrm{GV}$, which corresponds to the value that gives the best fit of antiproton secondaries to AMS-02 data.\footnote{We have checked that assuming a constant Fisk potential impacts only a limited range of intermediate DM masses, and, even in this mass range, leaving $\varphi$ free in the determination of the antiproton limits will impact our results only if a very large charge asymmetry in the solar modulation is present. In fact, we found that antideuteron fluxes above the experimental sensitivity could be obtained only by allowing $\varphi \geq 1.4\,\mathrm{GeV}$ (for reference coalescence momentum $p_0$; $\varphi \geq 1.2\,\mathrm{GeV}$ for largest $2\sigma$-allowed value for $p_0$). Such values are about double than the Fisk potential required to fit the proton spectrum.}

\begin{figure}[t]
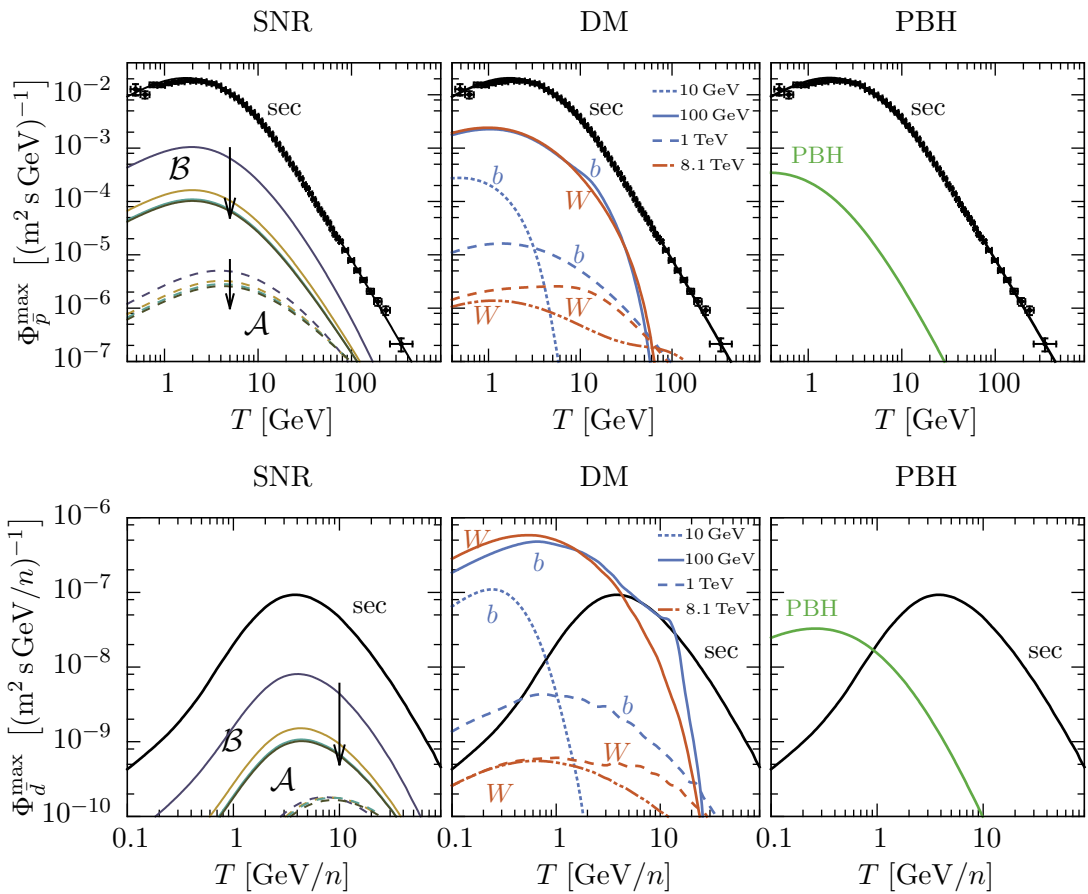

	\begin{center}
		\subimport{graphics/pbarlimiting/}{pbar3plot.tex} 
                \subimport{graphics/dbarLimiting/}{dbar3plot.tex}
		\caption{Maximum contribution to the antiproton flux allowed by current AMS-02 data (top panels), for the secondary contribution calculated in Section~\ref{subsubsec:secondary}, and maximal antideuteron flux at Earth (lower panels) from antibaryon production in SNRs (left panels), DM annihilation (central panels) and PBH evaporation (right panels). The arrows in the SNR fluxes indicate a growing value in the cutoff momentum $p_\mathrm{cut}=1$, $5$, $20$ and $100 \,\mathrm{TeV}$.}
		\label{fig-fluxes}
	\end{center}
\end{figure}

We show in Fig.~\ref{fig-fluxes}, top panels, the maximally allowed contributions to the antiproton flux from SNRs (left panel), DM annihilation (middle panel) and PBH evaporation (right panel) compatible with the AMS-02 data, for various choices of the free parameters in each of these sources. The values of the parameters that saturate the limits on the antiproton flux lead to the maximal injection rate of (anti-)baryons in the Galaxy, and therefore to the maximal antideuteron flux at Earth.  The corresponding maximal fluxes from each of these sources are shown in the bottom panels, confronted to the contribution expected from secondary production. For the decay of dark matter particles with mass $m_{\rm DM}$, not shown in the figures, the maximal fluxes can be approximated (up to a factor $\lesssim 5\%$ which stems from the different spatial distribution of the sources) by those obtained for the annihilation of DM particles with mass $m_{\rm DM}/2$ into the same final state.

More concretely, we consider for SNRs several values for the cutoff momentum $p_\mathrm{cut}=1$, $5$, $20$ and $100 \,\mathrm{TeV}$  and we consider the contribution to the antiproton flux only from the ${\cal A}$ term or only from the ${\cal B}$ term. As apparent from the plot, for higher values of $p_\mathrm{cut}$ the antiproton bounds get more stringent, and consequently the contribution to the antideuteron flux decreases. We find the maximum impact of the SNR contribution to the low energy antideuteron flux to be $\sim 10\%$ of the secondary flux for $p_\mathrm{cut}$ = 1 TeV and only ${\cal B}$-term contribution, while it is $< 5\%$ for larger $p_\mathrm{cut}$ values. We thus conclude that SNRs do not significantly enhance the total antideuteron flux. For DM annihilations we show $m_\mathrm{DM} = 10 \GeV$ (only for the $b\bar{b}$ final state), $100 \GeV$, $1 \TeV$ and $8.1 \TeV$ (only for the $W^+W^-$ case).  It follows from the central lower panel that annihilations of DM particles with mass below 100 GeV may enhance the total antideuteron flux by orders of magnitude. Finally, for PBH evaporation, the only free parameter is the local mass density. This scenario also offers good prospects for antideuteron detection and may also allow for a significant enhancement of the flux at Earth.

\begin{figure} [t]
	\begin{center}
		\subimport{graphics/dbarLimiting/}{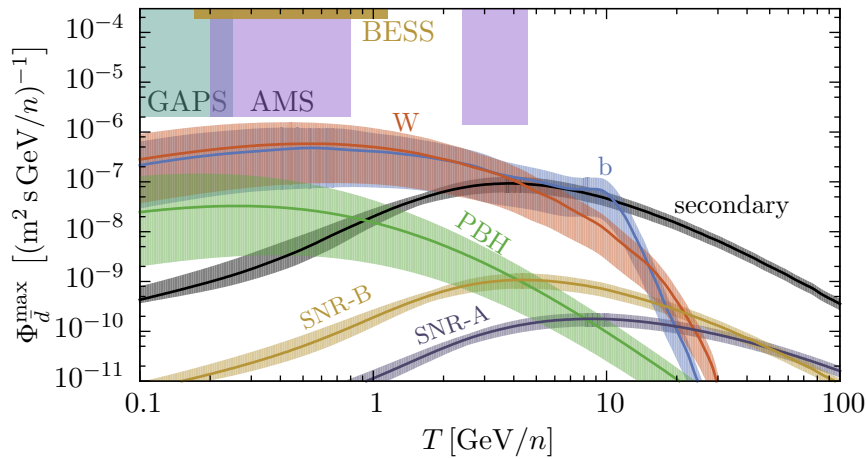}
		\caption{Maximal antideuteron flux from PBH evaporation, annihilation of DM particles with 80 GeV (102 GeV) mass into $\bar b b$ ($W^+W^-$), and production in SNRs assuming only ${\cal A}-$term or ${\cal B}-$term contributions and $p_{\rm cut}=20$ TeV, together with the expected flux from secondary production;  the bands bracket the uncertainty on the coalescence momentum and on solar modulation, as discussed in the text. For comparison we also show the current upper limit on the flux from BESS, as well as the projected sensitivities of AMS-02 and GAPS.}
	\label{fig-dbarLimiting}
	\end{center}
\end{figure}

We show in Fig.~\ref{fig-dbarLimiting} the maximal antideuteron fluxes allowed in each of these scenarios compatible with the current limits from the AMS-02 antiproton data, together with the expected flux from secondary production as well as the sensitivities of both GAPS and AMS-02 (the two energy windows for AMS-02 correspond to the time-of-flight (TOF) detector and the ring imaging Cherenkov (RICH) detector for the low-energy and high-energy windows, respectively). The shaded bands bracket the uncertainty related to the value of the coalescence momentum $p_0$ (which depends on the process, as detailed in Appendix~\ref{sec:production}) and of the Fisk potential of solar modulation (which we conservatively vary between 0.5 and 1.5 GV). Propagation uncertainties, on the other hand, have a small impact on the maximally allowed antideuteron fluxes, as argued in~\cite{Ibarra:2012cc,Fornengo:2013osa}, and are not included here. More concretely, the plot shows the maximal antideuteron flux from the annihilation of 80 GeV (102 GeV) DM particles into $b\bar b$ ($W^+ W^-$), from acceleration in SNRs assuming only ${\cal A}$-term or ${\cal B}$-term for the realistic value $p_{\rm cut}=20$ TeV, and from PBH evaporation. As apparent from the plot, the maximal antideuteron fluxes in all these scenarios are below the sensitivities of the GAPS and AMS-02 experiments, even for the most optimistic assumptions regarding coalescence momentum and solar modulation. 

\begin{figure} [t]
	\begin{center}
		\scalebox{1}{
		\subimport{graphics/limitingSignal/}{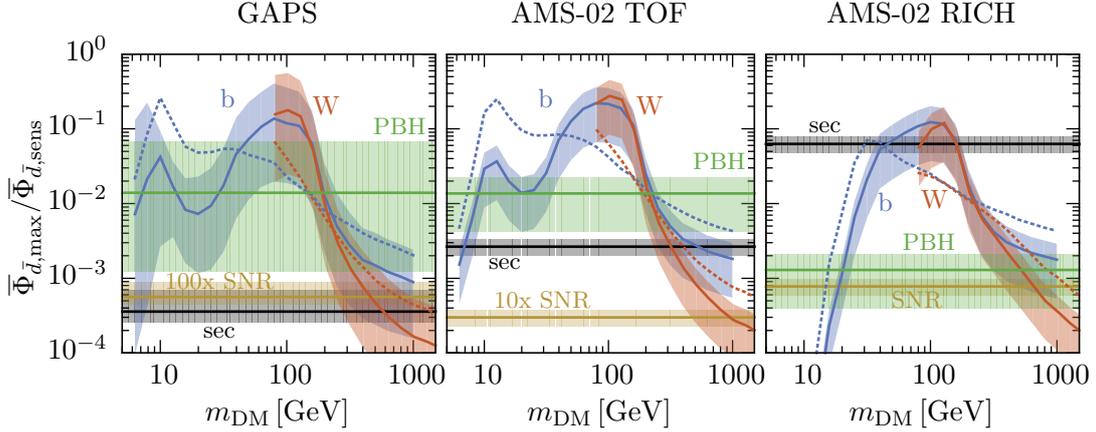}
		}
		\caption{Fraction of the sensitivities, as defined in the text, of GAPS (left panel), AMS-02 TOF (central panel) and AMS-02 RICH (right panel). Shown are contributions from PBH evaporation, DM annihilation into $\bar b b$ and $W^+ W^-$ (with mass ranging between 5 and 2000 GeV) and from production in SNRs considering only the ${\cal B}$-term contribution with $p_{\rm cut}=20$ TeV, as well as the expected contribution from secondary production. The dotted lines were determined assuming that the secondary flux exactly matches the AMS-02 $\bar p$ data.
For every contribution, solid lines represent the number of expected events obtained by assuming for $p_0$ the best fit values reported in Appendix \ref{sec:production} and a Fisk potential $\varphi = $ 0.9 GV.  As for Fig.~\ref{fig-dbarLimiting}, the bands bracket the uncertainty on the coalescence momentum and on solar modulation.
  }
	\label{fig-signal}
	\end{center}
\end{figure}

In order to make this statement more quantitative, we associate to the maximal allowed antideuteron fluxes a \emph{fraction of the experimental sensitivity} that is defined as follows: 
\begin{equation}
\mathcal{F} = \frac{\overline{\Phi}_{\bar{d},\mathrm{max}}}{\overline{\Phi}_{\bar{d},\mathrm{sens}}}= \frac{\int_{T_{\mathrm{min}}}^{T_{\mathrm{max}}} \mathrm{d}T\,\Phi_{\bar{d},\mathrm{max}}(T)}{\int_{T_{\mathrm{min}}}^{T_{\mathrm{max}}} \mathrm{d}T\,\Phi_{\bar{d},\mathrm{sens}}(T)}\;,
\end{equation}
where the integration is performed over the energy interval $[T_\text{min}, T_\text{max}]$ relevant to the considered experiment and $\Phi_{\bar{d},\mathrm{sens}}(T)$ is its sensitivity. The values that we adopt
for these quantities
are reported in Tab.~\ref{tab:exposure}.
The meaning of $\mathcal{F}$ is straightforward: if $\mathcal{F} \ge 1$, then the flux is equal to or above the projected experimental sensitivity, and discovery at the confidence level associated to the sensitivity ($99\%$ for GAPS, as reported in~\cite{Aramaki:2015laa}) is expected.
Furthermore, for experiments where the sensitivity is defined from the observation of a single antideuteron event, $\mathcal{F}$ would be equal to the number of events.
The results are shown in Fig.~\ref{fig-signal} for the GAPS (left panel), AMS-02 TOF (central panel) and AMS-02 RICH (right panel) detectors as a function of the DM mass.
In these plots, the contributions from secondary production, from SNRs and PBH evaporation are then constant (note also that for the GAPS and AMS-02 TOF detectors, the contribution from SNRs has been multiplied by a factor 100 and 10, respectively). As in Fig.~\ref{fig-dbarLimiting}, the shaded bands bracket the uncertainty related to the value of the coalescence momentum and the Fisk potential.

\begin{table}[t]
\center
\begin{tabular}{|c|c|c|c|}
\hline
Detector
& \begin{tabular}{c} $\Phi_{\bar{d},\mathrm{sens}}$ \\$[(\mathrm{m}^2\,\mathrm{s}\,\mathrm{GeV/ n})^{-1}]$ \end{tabular}
&  \begin{tabular}{c} $T_\mathrm{min}$\\$[\mathrm{GeV/n}]$ \end{tabular} 
&  \begin{tabular}{c} $T_\mathrm{max}$\\$[\mathrm{GeV/n}]$ \end{tabular} \\
\hline
GAPS~\cite{Aramaki:2015laa}& $2 \times 10^{-6}$  & $0.05$ & $0.25$ \\
\hline
AMS-02 TOF~\cite{Aramaki:2015pii} & $2 \times 10^{-6}$ & $0.2$ & $0.8$ \\
\hline
AMS-02 RICH~\cite{Aramaki:2015pii}& $1.4 \times 10^{-6}$ & $2.4$ & $4.6$ \\
\hline
\end{tabular}

\caption{Detector characteristics relevant for our analysis.}
\label{tab:exposure}
\end{table}

In all three detectors $\mathcal{F}$ is less than one and therefore detection is not expected. The scenario where $\mathcal{F}$ is found to be the largest is the annihilation of $80 \GeV$ DM particles into $b\bar b$ and the annihilation of $102 \GeV$ DM particles into $W^+ W^-$. In the latter scenario, and adopting for the coalescence momentum the best fit value $p_0= 192$ MeV and a Fisk potential $\varphi=0.90$ GV, the maximum  fraction of the sensitivity at GAPS is $ \mathcal {F} \simeq 0.2$  (which increases to $\simeq 0.3$ for $\varphi = 0.5 \;\mathrm{GV}$), while $\mathcal{F} \simeq 0.3$ and $\mathcal{F} \simeq 0.1$ at the AMS-02 TOF and RICH detectors, respectively (these values can be increased by a factor of $1.6$ for $p_0=226$ MeV, which is within $2\sigma$ in agreement with data, as discussed in Appendix~\ref{sec:production}).
It should be borne in mind, however, that this large antideuteron flux is the result of an excess in the AMS-02 antiproton data at around 10 GeV with respect to the background from secondary production (as calculated in Section~\ref{subsubsec:secondary}), and which favors the existence of an exotic component with a spectrum that resembles the one produced in these two scenarios. Should this excess be an artefact of a mismodeling of the secondary antiproton flux, then the expected antideuteron flux would be reduced.  This is also illustrated in Fig.~\ref{fig-signal} as dotted lines, which were computed assuming a hypothetical model of secondary production which exactly matches the AMS-02 $\bar{p}$ data, such that the only room for exotic contributions lies in the experimental uncertainties. In this hypothetical framework, the maximum antideuteron flux still reaches $\mathcal{F} \simeq 0.3$ in the low-energy range ({\it i.e.}~the energy intervals covered by GAPS and by the AMS-02 TOF detector), corresponding to the annihilation of 10 GeV DM particles into $b\bar{b}$.

An increase in the exposure would then be required in order to observe cosmic antideuterons. We find that, among the three detectors under consideration, the one with best prospects to observe the guaranteed flux from secondary production is the AMS-02 RICH detector. In this detector, on the other hand, the signal-to-background ratio for the exotic contributions is expected to be rather low.
In contrast, other detectors probing the low energy region of the antideuteron flux (GAPS and AMS-02 TOF) have better prospects to observe events from exotic sources.

\section{Prospects for antihelium}
\label{sec:antihe}

\begin{figure}[t]
	\begin{center}
		\subimport{graphics/He3Limiting/}{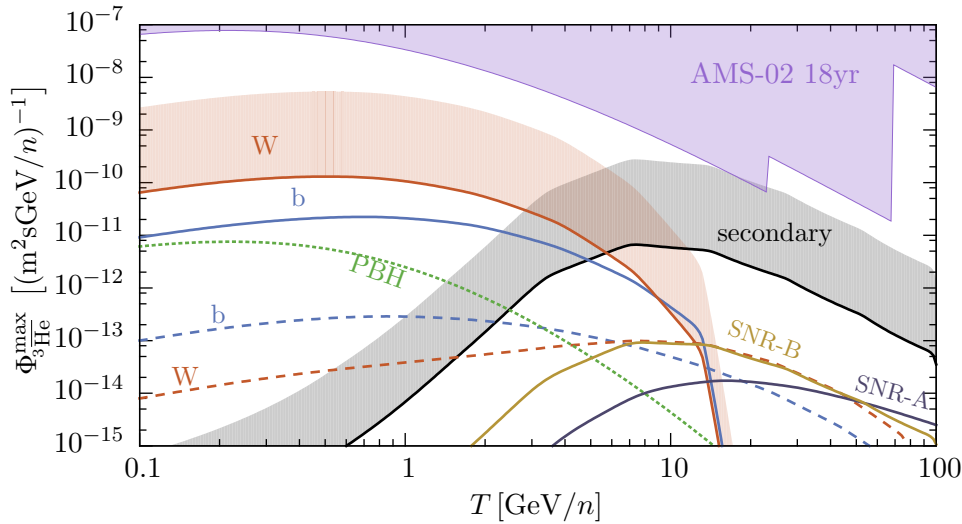}
		\caption{Maximal antihelium flux from PBH evaporation, annihilation of DM particles with $100 \GeV$ (line) or $1 \TeV$ (dashed) mass into $\bar b b$ or $W^+ W^-$, and production in SNRs assuming only $\mathcal{A}$-term or $\mathcal{B}$-term contributions and $p_\mathrm{cut} = 20 \TeV$, as well as the antihelium contribution from secondary production in the ISM. Lines correspond to $p_0^{\overline{\mathrm{He}}} = p_0^{\bar{d}}$, the shaded bands show the plausible range in coalescence momentum for the two most promising contributions (see text).
		}
	\label{fig-hebar}
	\end{center}
\end{figure}

The analysis that we carry out in this paper can be extended to heavier antinuclei. In particular, following Refs.~\cite{Carlson:2014ssa,Cirelli:2014qia}, one can consider the case of antihelium-3. The kinematics of spallation reactions is such that the energy threshold for the production of antinuclei increases with their mass number and therefore the antihelium flux of secondary origin is expected to be even more suppressed than in the antideuteron case. This translates into a potentially large enhancement of the total antihelium flux at low energies from exotic sources.

We compute the antihelium-3 flux produced by all the sources considered in this paper following the approach adopted in~\cite{Carlson:2014ssa} and assume that three antinucleons merge into an antihelium if they are within a minimum binding momentum-sphere of radius $p_0^{\overline{\mathrm{He}}}$. The formation of an antihelium can be realized either directly, with the coalescence of $\bar{p}\bar{p}\bar{n}$ or through the decay of an antitritium, which is formed through the coalescence of $\bar{p}\bar{n}\bar{n}$. We take into account only the latter channel, since for the former one the probability to form a bound state is expect to be suppressed because of the Coulombian repulsion between the two antiprotons. For $^3\overline{\rm He}$ production in nucleus-nucleus collisions and in DM annihilations or decays, we adopt the event-by-event coalescence model which is conceptually identical to the one that we have used for antideuterons, the only difference being that in this case the bound states that we are considering are composed of three antinucleons.  On the other hand, for production in PBH evaporation the event-by-event coalescence model is computationally unfeasible, hence we used instead the factorized coalescence model, which is intended as an order-of-magnitude estimate. In any case, when antinuclei are produced in a process with a small center of mass energy, as it is for PBH evaporation, we expect the factorized coalescence model to be fairly accurate, as shown for antideuterons in~\cite{Kadastik:2009ts,Fornengo:2013osa}. 

Experimental data that can be used to tune the coalescence model for the antihelium case are extremely scarce and refer only to nucleus-nucleus collisions. Therefore, we use values of the coalescence momentum in the wide range from 195 MeV to 357 MeV for DM annihilation and PBH evaporation, and from 167 MeV to 311 MeV for the secondary flux and for the contribution from SNRs. The lower limits of these intervals correspond to the coalescence momenta used for the antideuteron case, while the upper limits are obtained under the assumption that the coalescence momentum scales with the square root of the antinucleus binding energy~\cite{Carlson:2014ssa}: 
\begin{equation}
p_0^{\overline{\mathrm{He}}} = \left(\frac{B_{\overline{\mathrm{He}}}}{B_{\overline{\mathrm{d}}}}\right)^{1/2}\;p_0^{\bar{d}}\;.
\end{equation}
Galactic propagation and solar modulation are performed as detailed in Section~\ref{sec:propagation}, however the tertiary component of the fluxes has been neglected for antihelium, as the inelastic, non-destructive cross section for $^3{\rm He}$-$p$ collisions is not available. 
The cross sections for the interaction of antihelium with the interstellar gas are the ones used in~\cite{Carlson:2014ssa}. For the Fisk potential we set $\varphi = 0.9 \;\mathrm{GV}$. 

The maximum antihelium fluxes, compatible with the current AMS-02 antiproton data, for all the sources under scrutiny are shown in Fig.~\ref{fig-hebar}, confronted to the projected AMS-02 sensitivity after 18 years of data taking~\cite{2010arXiv1009.5349K}. For the case of DM annihilation, we consider two masses: 100 GeV (solid lines) and 1 TeV (dashed lines). The conclusions are very similar to the ones drawn from  Fig.~\ref{fig-dbarLimiting} for antideuterons: at energies below a few GeV/n, the exotic contributions may significantly enhance the total antihelium flux, the potential enhancement being largest for DM annihilations and most modest for SNRs. For kinetic energies per nucleon larger than $\sim 10$ GeV, on the other hand, these exotic sources only give a subdominant contribution to the flux. Therefore, experiments sensitive to these energies will only be able to probe secondary antihelium production, but are unlikely to probe the exotic mechanisms of antimatter production discussed in this paper. It is interesting to remark that for $p_0^{\overline{\mathrm{He}}} $ = 311 MeV the detection of cosmic antihelium may be within the reach of AMS-02, provided data can be collected over at least 18 years.

\section{Conclusions}
\label{sec:conclusions}

We have presented a comprehensive analysis of the cosmic antinuclei signals at Earth from all known and hypothetical antibaryon sources in our Galaxy: collisions of high energy cosmic rays with the interstellar gas, production in old supernova remnants, primordial black hole evaporation and dark matter annihilation or decay. The good agreement of the predictions from secondary production with the AMS-02 antiproton data severely restricts the rate of antibaryon injection in the interstellar medium from supernova remnants and from exotic sources. We find that supernova remnants can only contribute  to at most $10\%$ of the total antideuteron flux at the energies relevant for the experiments AMS-02 and GAPS. On the other hand, both dark matter annihilation/decay and primordial black hole evaporation can generate a large excess of low-energy antideuterons over the astrophysical background. This result reinforces the role of low-energy antideuterons as a golden channel for the discovery of exotic processes.

Our analysis indicates that the very large signal-to-background ratios that we find for the exotic processes studied in this paper do not translate into a visible signal for the present generation of experiments.
In fact, we find that, for typical values of the coalescence momentum, the detection of cosmic antideuterons will require an increase of the experimental sensitivity compared to ongoing and planned instruments by at least a factor of 2.
It should be born in mind, though, that the derivation of this theoretical maximal flux involved a number of assumptions, therefore, given that the maximal flux is of the same order of magnitude as projected sensitivities, a discovery of an antideuteron component in cosmic rays in the near future ought not to be precluded. Concretely, larger antideuteron signals can be possible if {\it i)} the actual antideuteron production rate is larger than the one predicted by the model employed in this work, either because the actual coalescence momentum deviates more than $2\sigma$ from the central value assumed in this analysis, or simply because the coalescence model does not correctly describe the processes relevant for antideuteron production in our Galaxy; {\it ii)}  the antiproton flux receives sizable contributions from a mechanism other than secondary production in the ISM, such that the upper limit on the dark matter annihilation cross section, density of primordial black holes, or normalization factors of the $\mathcal{A}$- and $\mathcal{B}$-terms get relaxed, thus allowing a larger antideuteron signal; in the case of dark matter, however, one should consider that such a scenario could be in tension with the constraints that are derived from other indirect detection channels, e.g.~gamma-rays {\it iii)} there are additional production mechanisms, not considered in this work, with a larger antideuteron yield.


We have also investigated the features of the antihelium fluxes produced by all the aforementioned sources, again in connection with the bounds derived from AMS-02 antiproton measurements. Our results show that also in the antihelium case, we expect to have large signal-to-background ratios in the low-energy window for both dark matter annihilation/decay and primordial black hole evaporation. On the other hand, the flux produced by supernova remnants is expected to be well below the spallation background. Whereas AMS-02 may observe, under very optimistic conditions, cosmic antihelium nuclei from secondary production, the prospects for detecting signals from exotic sources in the antihelium flux are much poorer than in the antideuteron flux.

\vspace{-1ex}

\paragraph{Acknowledgements}

This work has been partially supported by the DFG cluster of excellence Origin and Structure of the Universe, the German-Israeli Foundation for Scientific Research, the TUM Graduate School and the Studienstiftung des Deutschen Volkes. We are grateful to Eric Carlson for providing the $^3\overline{\rm He}$ multiplicities from dark matter annihilation/decay and to Michael Kachelrie\ss\ for discussions about particle acceleration in supernova remnants.

\vspace{-1ex}

\appendix
\section{The coalescence model for antideuteron production}
\label{sec:production}
\vspace{-2ex}

In this Appendix, we briefly review the coalescence model, which we employ for calculating the injection spectrum of antideuterons produced by the hadronization process of quarks and gluons. In this approach, one assumes that a $\bar p \bar n$ pair forms an antideuteron if the relative momentum $\Delta p$ of the antinucleons is smaller than the coalescence momentum $p_0$~\cite{Schwarzschild:1963zz}, and if in addition the antinucleons are produced within a relative spatial distance of $\simeq 1\,$fm~\cite{Ibarra:2012cc,Fornengo:2013osa}. The coalescence momentum is a free parameter of the model which must be determined from experimental data. 

More precisely, in this work we employ the event-by-event coalescence model~\cite{Kadastik:2009ts} using an appropriate Monte Carlo event generator both for the determination of $p_0$ as well as for the prediction of the spectrum of antideuterons produced in the various processes of interest. For DM annihilations and PBH evaporation we use PYTHIA 8.176~\cite{Sjostrand:2007gs}, with the coalescence momentum tuned to the data from ALEPH~\cite{Schael:2006fd}, which measured the total antideuteron yield in the decay of the $Z$ boson, as well as from BaBaR~\cite{Lees:2014iub}, which observed antideuteron production in $e^+ e^- \rightarrow q \bar q$ collisions at $\sqrt{s}=10.58\,$GeV. The envelope of the corresponding $2\sigma$-preferred regions for $p_0$ obtained by a fit of PYTHIA to the ALEPH and \textsc{BaBar} data is 128.7 -- 226.1 MeV~\cite{Ibarra:2012cc,Aramaki:2015pii}, which we will use as the range of possible coalescence momenta for DM annihilations and PBH evaporation. As a benchmark choice we use $p_0 = 192\,$MeV, corresponding to the central value of the ALEPH measurement. On the other hand, for $pp$ and $\bar p p$ collisions relevant for the spallation of CRs on the interstellar matter as well as for antideuteron production in SNR, we employ the antideuteron spectra obtained in~\cite{Ibarra:2013qt}, which are based on an implementation of the event-by-event coalescence model in the Monte Carlo generator DPMJET-III~\cite{Roesler:2000he}. In this case, we consider values of $p_0$ in the range of 138.2 -- 163.8 MeV~\cite{Ibarra:2013qt,Aramaki:2015pii}, corresponding to the $2\sigma$-allowed region obtained by a fit to the CERN ISR data on $pp$ collisions at $\sqrt{s} = 53\,$GeV~\cite{Alper:1973my,Henning:1977mt}. Our benchmark results for the antideuteron flux from CR spallations and from SNR will be based on the central value of the fit, $p_0 = 152\,$MeV. For a more detailed discussion on the uncertainties related to the coalescence model we refer to~\cite{Ibarra:2012cc,Aramaki:2015pii,Dal:2012my,Dal:2014nda,Dal:2015sha}.

\vspace{-2ex}

\bibliographystyle{JHEP}
\bibliography{antideuteron}

\end{document}